\documentclass[aps,prl,twocolumn,showpacs,floatfix,superscriptaddress]{revtex4}
\usepackage{graphicx}
\newcommand {\be}{\begin{equation}}
\newcommand {\ee}{\end{equation}}

\begin{document}

\title{ Self--Consistent Mode--Coupling Approach to 1D Heat Transport}
\date{\today}

\author{Luca Delfini}
\affiliation{Istituto Nazionale di Ottica Applicata, largo E. Fermi 6
I-50125 Firenze, Italy}

\author{Stefano Lepri}
\email{stefano.lepri@isc.cnr.it}
\affiliation{Istituto dei Sistemi Complessi, Consiglio Nazionale delle
Ricerche, largo E. Fermi 6
I-50125 Firenze, Italy}

\author{Roberto Livi}
\altaffiliation[Also at ]{Istituto Nazionale di Fisica Nucleare 
and Istituto Nazionale di Fisica della Materia, Firenze.}
\affiliation{Dipartimento di Fisica, via G. Sansone 1 I-50019, Sesto
Fiorentino, Italy }

\author{Antonio Politi}
\affiliation{Istituto dei Sistemi Complessi, Consiglio Nazionale delle
Ricerche, largo E. Fermi 6
I-50125 Firenze, Italy}

\begin{abstract}

In the present Letter we present an analytical and numerical solution of the
self-consistent mode-coupling equations for the problem of heat conductivity in
one-dimensional systems. Such a solution leads us to propose a different
scenario to accomodate the known results obtained so far for this problem.
More precisely, we conjecture that the universality class is determined by
the leading order of the nonlinear interaction potential. Moreover, our analysis
allows us determining the memory kernel, whose expression puts on a
more firm basis the previously conjectured connection between anomalous 
heat conductivity and anomalous diffusion.
\end{abstract}


\maketitle

It is well known that relaxation and transport phenomena in reduced spatial
dimensions ($d<3$) are often qualitatively different from their
three-dimensional counterparts. This is a documented effect, for
example, in single-filing systems, where particle diffusion does not follow
Fick's law~\cite{singlef}. Another related phenomenon is the enhancement
of  vibrational energy transmission in quasi-$1D$ systems like
polymers~\cite{morelli} or individual carbon nanotubes~\cite{nanotube}.
The specific instance of anomalous thermal conduction in low-dimensional
many-particle systems has recently received a renewed attention \cite{LLP97}.
Anomalous behaviour means both a divergence of the finite-size conductivity 
$\kappa(L)\propto L^\alpha$ in the large--size limit $L\to \infty$ and  a
nonintegrable decay of equilibrium correlations of the energy current
(the Green-Kubo integrand), $\langle J(t)J(0)\rangle \propto t^{-(1-\alpha)}$
at large times $t\to \infty$ (with $0\le\alpha < 1$). Simulations ~\cite{rep}
and theoretical arguments \cite{NR02} indicate that anomalies should occur
generically in $d\le 2$ whenever momentum is conserved.

The importance of predicting the scaling behaviour (i.e. the value of $\alpha$)
is twofold: (i) on a basic ground, to classify the ingredients (e.g.
symmetries) that define the possible universality classes; (ii) on a practical
ground, to estimate heat conductivity in finite systems, a crucial issue to
compare with experimental data on, say, carbon nanotubes. In spite of several
efforts,  the theoretical scenario is still controversial. In $d=1$, arguments
based on  Mode-Coupling Theory (MCT) \cite{E91,L98}, a well-known approach to
estimate long-time tails of fluids \cite{PR75} and to describe the glass
transition \cite{mct}, yield  $\alpha = 2/5$. This estimate was criticized as
inconsistent in Ref.~\cite{NR02}, where renormalization group arguments were
instead shown to give $\alpha =1/3$. Nevertheless, the 2/5 value has been
later derived both from a kinetic-theory calculation for the quartic ($\beta$)
Fermi Pasta Ulam (FPU) model \cite{P03} and from a solution of the MCT by
means of an {\it ad hoc} Ansatz \cite{li}. It was thereby conjectured \cite{li}
that 2/5 is found for a purely longitudinal dynamics, while a
crossover towards 1/3 can to be observed only in the presence of a coupling to
transversal motion. Unfortunately, the accuracy of numerical simulations is
generally insufficient to disentangle the whole picture. The only two
convincing studies concern the hard point gas, which has been recently found to
be characterized by $\alpha = 1/3$ \cite{denis}, and the purely quartic FPU
model, where instead $\alpha$ is definitely larger than 1/3 (and possibly
closer to 2/5) \cite{LLP03}. The situation is even more controversial in $d=2$
where logarithmic divergence is expected \cite{2d}.

The exact self-consistent solution of the MCT equations presented in this
Letter demonstrates that the overall scenario is different from that one
proposed in \cite{li}, namely that $\alpha =1/3$ in the presence of cubic  
nonlinearities.  This prediction is confirmed by our numerical simulations of 
the FPU model with cubic potential which yields sizably different $\alpha$
values with respect to the quartic case. Altogether, theoretical and numerical
results indicate that the asymptotic scaling behaviour is determined by the
order of the leading nonlinearity in the interaction potential.

Let us consider the simplest one-dimensional version of the self-consistent
MCT for the normalized correlator of the Fourier transform
of the displacement field $G(q,t)= \langle Q^*(q,t)Q(q,0) \rangle/\langle
|Q(q)|^2 \rangle$. In dimensionless units in which the particle mass, the
lattice spacing and the bare sound velocity are set to unity, 
they read ~\cite{SS97,L98}
\begin{eqnarray}\nonumber
&&\hskip -0.5cm{\ddot G} (q,t) + 
\varepsilon \int_0^t \Gamma (q,t-s) {\dot G}(q,s) \, ds 
+ {\omega}^2(q) G(q,t)  
= 0 \\
&&\hskip -0.5cm \Gamma(q,t)= \,\omega^{2}(q)
\,\frac{2 \pi}{N} \sum_{p+p'-q=0,\pm\pi}  \,G(p,t) G(p',t) \quad .
\label{mct}
\end{eqnarray}
We consider periodic boundaries so that the wavenumbers are $q=2\pi k/N$ with
$-N/2+1 \leq k \leq N/2$. Notice that
$G(q,t)=G(-q,t)$. Eqs.~(\ref{mct}) must be solved with the initial conditions
$G(q,0)=1$  and $\dot G(q,0)=0$. 

The first of Eqs.~(\ref{mct}) is exact and is derived within the 
well--known Mori--Zwanzig projection approach \cite{KT}. In the 
small--wavenumber limit, it describes the response of an elastic string at
finite temperature. The above mode-coupling approximation
of the memory function $\Gamma$ has been derived for a chain of atoms
interacting through a nearest-neighbour anharmonic potential $V(x)$
\cite{SS97,L98} whose expansion around its minimum at $x=0$ is of the form
$x^2/2+g_3x^3/3+\ldots$ (see, e.g., the Lennard-Jones potential). Both the
coupling constant $\varepsilon$ and the dispersion relation $\omega(q)$ are
temperature-dependent input parameters that must be computed independently by
simulation or approximate analytical approaches \cite{SS97,L98}. For the aims
of the present Letter, we may restrict ourselves to considering their bare
values, obtained in the harmonic approximation that, in our units, read
$\varepsilon = {3g_3^2 k_BT / 2\pi}$ and $\omega(q)=2 | \sin{q\over 2}|$. Of
course, the actual renormalized values are needed when a quantitative
comparison with a specific model is looked for. Moreover, since the anomalies
we are interested in stem from the nonlinear interaction of long-wavelength
modes, we let $\omega(q)=|q|$, in the analytic treatment
presented below. 

Direct numerical simulations \cite{L98} indicate that nonlinear and nonlocal
losses in Eq.~(\ref{mct}) are small compared to the oscillatory terms. This
suggests splitting the $G$ dynamics into phase and amplitude evolution,
\begin{equation}
G(q,t) \;=\; C(q,t )e^{i \omega (q)t} + c.c.
\label{g}
\end{equation}
Upon substituting this equation into Eq.~(\ref{mct}), one obtains, in the slowly
varying envelope approximation, $qC \gg \dot{C}$,
\begin{equation}
2\frac{\partial}{\partial t}C(q,t ) + 
\varepsilon \int_0^{t}\, dt ' \, M(q,t-t ') C(q,t ') \;=\; 0
\label{eqc}
\end{equation}
plus a similar expression for $C^*$, while the new kernel $M$ turns out to be
\begin{eqnarray}
M(q,t) = q^2 \int\limits_{-\infty}^{\infty}dp \, C^{\ast}(p-q,t)C(p,t )\quad ,
\label{eqm}
\end{eqnarray}
where the sum in (\ref{mct}) has been replaced by a suitable integral, since
we consider the thermodynamic limit $N=\infty$ and 
small $q$-values, which are, by the way, responsible for the
asymptotic behavior.

Notice that Eqs.~(\ref{eqc},\ref{eqm}) have been obtained after discarding the
second order time derivative of $C(q,t)$ as well as the integral term
proportional to $\dot C$, besides all rapidly rotating terms. The validity
of this approximation is related to the separation between the decay rate of
$C(q,t)$ and $\omega(q)$; its correctness will be cheked a posteriori, after
discussing the scaling behaviour of $C(q,t)$. Notice also that in this
approximation, Umklapp processes do not contribute: it is in fact well
known that they are negligible for long--wavelength phonons in 1D ~\cite{rep}. 

Having transformed the second order differential equation for $G$ into a first
order one for $C$, we can introduce a simple scaling argument yielding the
dependence of $C$ on $q$ and $t$ as follows
(see also \cite{LB94}, where a similar equation was investigated),
\begin{equation}
C(q,t) = g(\sqrt{\varepsilon}t q^{3/2}) \quad,\quad   
M(q,t) = q^{3}f(\sqrt{\varepsilon}t q^{3/2}) .
\label{gf3}
\end{equation}
This shows that the decay rate for the evolution of $C(q,t)$ is of the order 
$q^{3/2} \sqrt{\varepsilon}$, which has to be compared with the scale $q$ of the
corresponding phase factor. Accordingly, amplitude and phase dynamics become
increasingly separated for $q \to 0$. High $q$-values ($q \approx 1$) are those
for which the slowly varying envelope approximation is less accurate. However,
if $\varepsilon$ is small enough, such modes are correctly described, too.

\begin{figure}
\begin{center}
\includegraphics[clip,width=7.5cm]{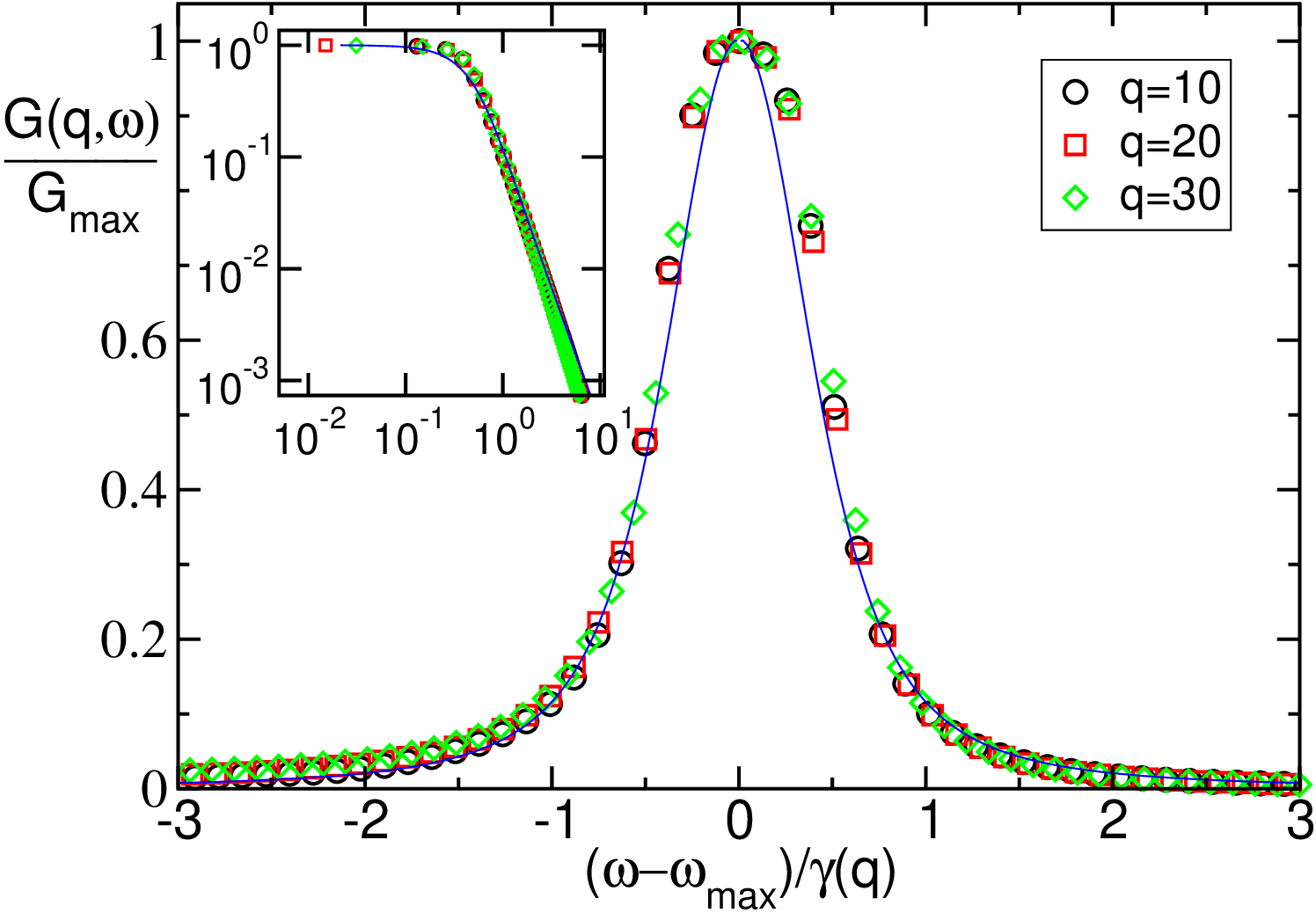}
\caption{Fourier transform $G(q,\omega)$ of the correlation functions for 3
different wavenumbers $\varepsilon=1$, $N=2000$. The solid line is the
lineshape computed by the approximate analytic theory, i.e. by Fourier
transforming the function $C(q,t)$ defined in Eqs.~(\ref{gf3},\ref{smallx}).
The same curves are plotted in the inset in log-log scales, where only positive
frequencies are shown.}
\label{spettri}
\end{center}
\end{figure}

\begin{figure}
\begin{center}
\includegraphics[clip,width=7.5cm]{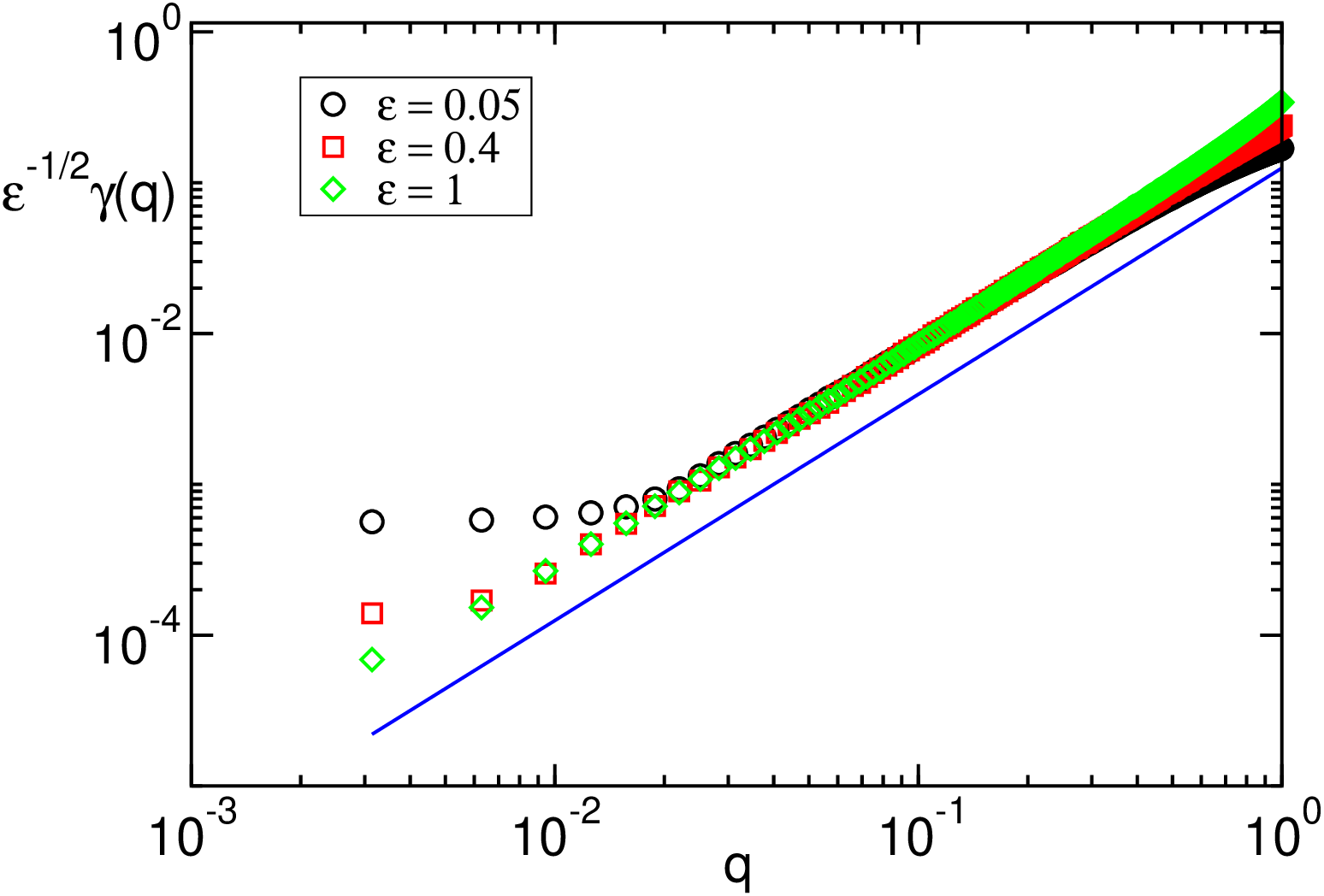}
\caption{Scaling of the linewidth $\gamma (q)$ of $G(q,\omega)$ with $q$ 
for 3 different values of the coupling constant $\varepsilon$ and $N=2000$.
The solid line corresponding to the power law $q^{3/2}$ is plotted for
reference.}
\label{gamma2}
\end{center}
\end{figure}

The functions $f$ and $g$ can be determined by substituting the 
previous expressions into Eqs.~(\ref{eqc},\ref{eqm}). Upon setting
$x=\sqrt{\varepsilon}t q^{3/2}$, one obtains the equation
\begin{eqnarray}
&&g'(x) = -\int_0^x dy \, f(x-y) g(y) \\
\label{gx}
&&f(x) = x^{-2/3}\int_{-\infty}^{+\infty} dy
g^\ast (\mid x^{2/3}-y\mid ^{3/2}) g(y^{3/2})  
\label{fx}
\end{eqnarray}
The asymptotic behaviour for $x\to 0$ can be determined analytically,
\begin{equation}
g(x) \;=\; \frac12 \exp \Big(-{a x^{4/3}\over 4} \Big) \quad,\quad 
f(x) \;=\; \frac{a}{x^{2/3}}
\label{smallx}
\end{equation}
where $a$ is a suitable constant which is determined self-consistently from 
Eq.~(\ref{gx}).
To assess the validity of the above calculation, we have numerically integrated
Eqs.~(\ref{mct}) by the Euler method for the original dispersion relation
$\omega(q)$ and different $\varepsilon$-values. We have verified that a time
step $\Delta t = 0.01$ guarantees a good numerical accuracy over the
explored time range. The Fourier transform $G(q,\omega)$ is plotted
in Fig.~\ref{spettri} for three different $q$-values versus
$\omega - \omega_{max}(q)$, where $\omega_{max}(q)$ is the frequency
corresponding to the maximal value $G_{max}$ of the spectrum (this is 
equivalent to removing the oscillating component from $G(q,t)$). 
Furthermore, in order to test relation (\ref{gf3}),
the vertical axis is scaled to the maximum $G$-value, while the
frequencies  are divided by the half--width $\gamma(q)$ at half of the maximum
height. This latter quantity can be interpreted as the inverse lifetime of 
fluctuations of wavenumber $q$. The good data collapse confirms the existence of
a scaling regime. The approximate analytical expression is in excellent agreement
with the data. As expected, some deviations are present for small $\omega$ 
where Eq. ~(\ref{smallx}) is not strictly applicable. Moreover, in the
inset of Fig.~\ref{spettri}, where the same curves are plotted using doubly
logarithmic scales, one sees that the lineshapes follow the predicted
power-law, $\omega^{-7/3}$, over a wide range of frequencies. In
Fig.~\ref{gamma2} we show that $\gamma(q)$ is indeed proportional to
$\sqrt{\varepsilon}\, q^{3/2}$. It is particularly instructive to notice 
that the agreement is very good also also for a relatively large value of
the coupling constant ($\varepsilon\simeq 1$), although the slowly varying
envelope approximation is not correct for large $q$ values. The deviations
observed at small $q$-values for small couplings are due to
the very slow convergence in time. Better performances could be 
obtained by increasing both $N$ then the integration time 
($10^4$, in our units) well beyond our current capabilities.

It is crucial to compare these result with the previous work. Making use of
Eqs.~(\ref{gf3},\ref{smallx}), it can be shown that the memory function
$\Gamma$ contains terms of the form $ q^2 \, e^{\pm iqt}/t^{2/3}$, i.e. it
oscillates with a power--law envelope. Accordingly, its Laplace trasform has 
branch-cut singularities of the form  $q^2/(z \pm q )^{\frac{1}{3}}$. This
finding is not consistent with the heuristic assumption of Refs.~\cite{E91,L98}
and the result of \cite{li}, where MCT equations 
were solved with
the Ansatz $\Gamma(q,z)=q^2\mathcal{V}(z)$. In addition, the numerical solution does 
not show any signature of the $q^2/z^{1/3}$ dependence found in ~\cite{li}. 
For instance, it would imply a peak at $\omega=0$ in the spectrum of
$\Gamma$ which is, instead, absent in numerical solutions. 

In order to estimate the long-time decay of the energy--current 
autocorrelation we use the approximate expression proposed in \cite{li},
\be
\langle J(t)J(0) \rangle \; \propto \; 
\sum_{q} \, v^2(q) G^2(q,t) 
\label{flusso}
\ee 
where $v(q)=\omega'(q)$ is the bare group velocity. Neglecting the oscillating
terms in $G^2$ and letting $v(q)\simeq 1$ (the sum is dominated by the
small--$q$ terms), we can use Eq.~(\ref{gf3}) to find
\be
\langle J(t)J(0)\rangle \;\propto \; \int dq \, g^2(\sqrt{\varepsilon} t q^{3/2})
\;\propto\; \frac{1}{t^{2/3}}
\ee
i.e. $\alpha=1/3$. Note that this result is independent of the actual form of
$g$ (provided convergence of the integral is insured). This scaling has also
been checked to hold by directly evaluating the sum (\ref{flusso}) with 
the numerically computed correlations $G$. 
\begin{center}
\begin{figure}
\includegraphics[clip,width=7.5cm]{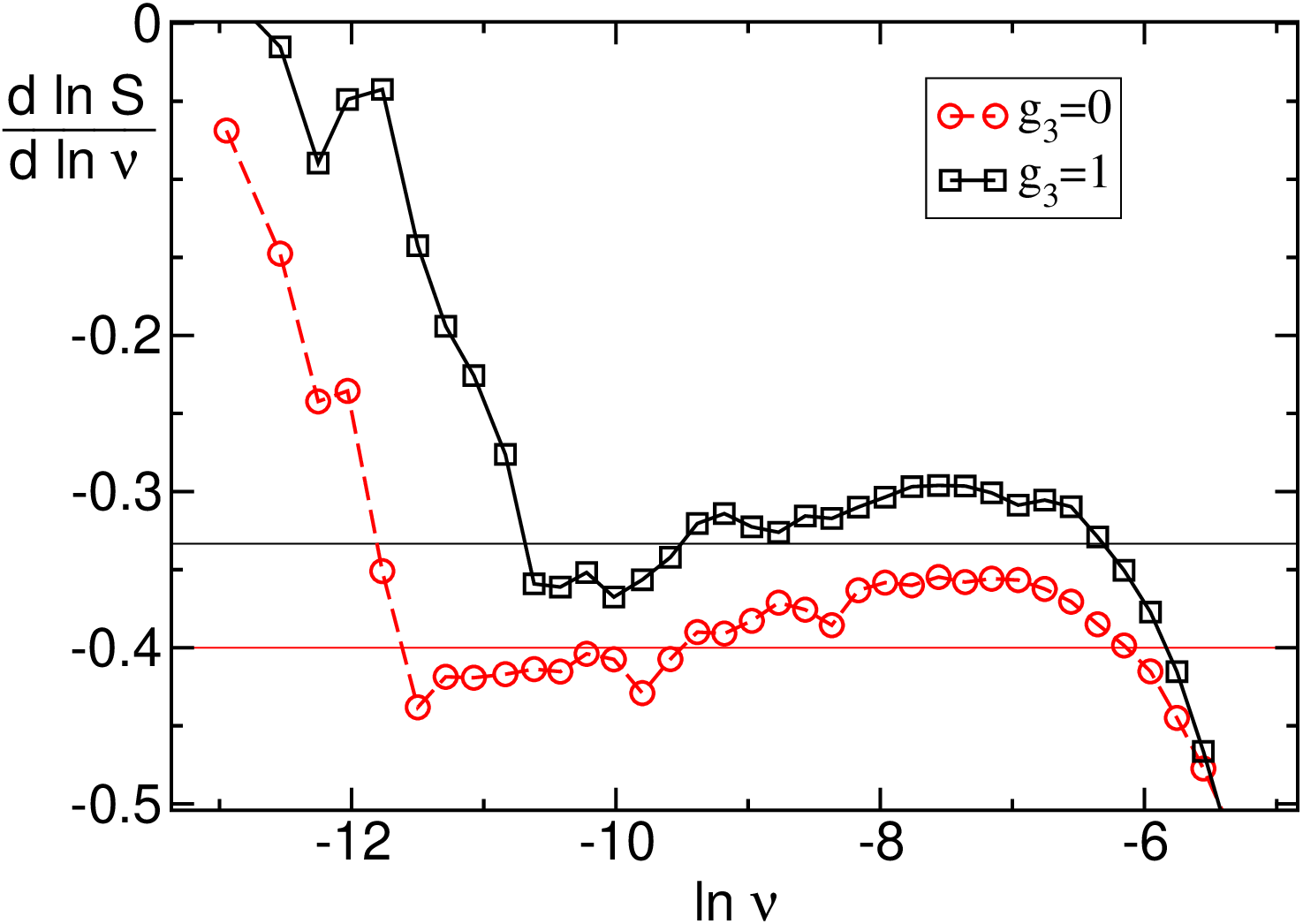}
\caption{ 
The logarithmic derivative of the energy-flux spectrum $S(\nu)$ versus the
frequency $\nu$ for the FPU potential with energy density set equal to 10. The
two horizontal lines correspond to the theoretical predictions $-1/3$ and $-2/5$.
The statistical error is on the order of the observed irregular fluctuations.}
\label{fig:deriv}
\end{figure}
\end{center}
\vspace{-0.9cm}
As already mentioned at the beginning, several numerical simulations have been
performed to estimate the exponent $\alpha$. Since the most accurate data
\cite{LLP03} differ significantly from 1/3, we decided to run a new set of
simulations. We considered the FPU model with interparticle
potential $V(x) = x^2/2 + g_3 x^3/3 + x^4/4$ and periodic boundary conditions.
We performed microcanonical simulations to compute the average power
spectrum $S(\nu)$ of $J$ (see Refs.~\cite{rep,LLP03} for details). The long--time
tail manifests itself as a power--law divergence $\nu^{-\alpha}$ at
low-frequencies, $\nu \to 0$. In order to provide a reliable estimate of
$\alpha$, it is convenient to evaluate the logarithmic derivative
${d \ln S}/{d \ln \nu}$. As shown in Fig.~\ref{fig:deriv}, for $g_3=1$ this
quantity does display a plateau around $-1/3$ (the growth towards zero at very
small $\nu$ values is due to the cutoff introduced by the finite size of the
lattice). On the other hand, the data obtained for $g_3=0$ (FPU-$\beta$ model)
indicate a sensibly different scaling exponent, which is much closer to $-2/5$
and in agreement with previous works ~\cite{P03,LLP03}.

We have thus reached the important conclusion that the memory kernel decays 
algebraically in the prescribed regime and, accordingly, the relaxation 
is not exponential $\log C \simeq - q^2 t^{4/3}$ (i.e., non Lorenzian
lineshapes). The further striking feature is that conventional hydrodynamics 
breaks down, since the peak widths scale as $q^{3/2}$ rather
than $q^2$, as expected in the standard case.  In addition, the linewidths are
connected to transport coefficients being proportional to $\Lambda q^2$, where
$\Lambda$ is the sound attenuation constant. The anomalous scaling can be
recasted in terms of a diverging $\Lambda(q) \sim q^{-1/2}$. 
Altogether, one may think of this as a superdiffusive process, intermediate 
between standard diffusive and ballistic propagation. Our result thus strenghten
the picture emerging in Ref.~\cite{denis} from the analysis of the hard--point gas,
where it has been shown that energy perturbations perform a Levy walk. One
merit of our approach is that it allows for a direct connection with anomalous
diffusion problems \cite{adiff}. As it is known, these can be modeled by
generalized Langevin equations with power-law kernels. If we now assume that
expression (\ref{smallx}) for $f$ holds for every $x$, we can solve
Eqs.~(\ref{eqc}) by Laplace transforming Eq.~(\ref{eqm}) to obtain 
$C(q,z) = iz^{1/3}/(iz^{4/3}+ a q^{2})$.
This expression is precisely the Laplace transform of the Mittag-Leffler
function $E_{\mu}(-(\lambda t)^{\mu})$~\cite{maina,adiff} for $\mu = 4/3$ and
$\lambda= (a q^2)^{3/4}$ \cite{nota}.
This observation suggests that the effective evolution of 
fluctuations should be modeled by the fractional differential equation
\be
{\partial ^\mu \over\partial t ^\mu} C(q,t) + \lambda^\mu  C(q,t) 
\;=\; 0 .
\ee
The case of interest here ($1<\mu\le 2$) corresponds to the so--called fractional 
oscillations \cite{maina}. It should be emphasized that in the present context,
memory arises as a genuine many-body effect and needs not to be postulated 
{\it a priori}.

In conclusion, we have shown that MCT with cubic nonlinearity (\ref{mct})
predicts a $t^{-2/3}$ decay of the heat current autocorrelation, i.e.
$\alpha=1/3$. Our analysis reconciles this approach with the
renormalization-group calculation \cite{NR02} and supports the idea that the
mechanisms yielding anomalous transport in 1D are largely universal. The
sizeable deviations observed for quartic potentials suggest the existence of a
different universality class which should be described by different
mode-coupling equations with a quartic nonlinearity. A preliminary analysis 
confirms that this scenario is indeed correct \cite{futuro}. Finally, we have
analytically shown that memory effects emerging from the nonlinear interaction
of long-wavelength modes can be described by a generalized Langevin equation
with power-law memory. This provides a sound basis establishing a connection
between anomalous transport and superdiffusive processes.  

We acknowledge useful discussions with A. Cuccoli and H.~Van Beijeren. This work
is partially supported by the PRIN2003 project {\it Order and chaos in
nonlinear extended systems} funded by MIUR-Italy.


\begin{thebibliography}{00}

\bibitem{singlef} K. Hahn, J. K\"arger, and V. Kukla, 
Phys. Rev. Lett. {\bf 76}, 2762 (1996).

\bibitem{morelli} D.T. Morelli \textit{et al.}, 
Phys. Rev. Lett. {\bf 57}, 869 (1986).

\bibitem{nanotube} S. Berber, Y. Kwon, D. Tomanek, Phys. Rev. Lett.
{\bf 84}, 4613 (2000); J. Hone \textit{et al.}, 
Phys. Rev. B {\bf 59} R2514 (1999);
S. Maruyama, Physica B {\bf 323}, 193 (2002). 
N. Mingo and D. A. Broido, Nanoletters {\bf 5} 1221 (2005).

\bibitem{LLP97} S. Lepri, R. Livi, A. Politi, Phys. Rev. Lett. {\bf 78}, 1896
(1997); Europhys. Lett. {\bf 43}, 271 (1998). 

\bibitem{rep} S. Lepri, R. Livi, A. Politi, Phys. Rep. {\bf 377}, 1 (2003).

\bibitem{NR02}  O. Narayan, S. Ramaswamy, Phys. Rev. Lett.
{\bf 89},  200601 (2002).

\bibitem{E91} M.H. Ernst, Physica D {\bf 47}, 198 (1991).

\bibitem{L98} S. Lepri, Phys. Rev. E {\bf 58} 7165 (1998).


\bibitem{PR75} Y. Pomeau, R. R\'esibois, Phys. Rep. \textbf{19},  63 (1975).

\bibitem{mct} W. G\"otze in 
\textit{Liquids, freezing and the glass transition},
edited by J. P. Hansen, D. Levesque and J. Zinn-Justin 
(North Holland, Amsterdam 1991); R. Schilling in 
\textit{Collective dynamics of nonlinear and 
disordered systems}, edited by G.Radons, W. Just and P. H\"aussler 
(Springer, Berlin, 2003).

\bibitem{P03} A. Pereverzev,
Phys. Rev. E \textbf{ 68}, 056124 (2003)

\bibitem{li} J-S. Wang and B. Li, Phys. Rev. Lett. {\bf 92} 074302 (2004);
Phys. Rev. E {\bf 70} 021204 (2004).

\bibitem{denis} P. Cipriani, S. Denisov and A. Politi, Phys. Rev. Lett.
{\bf 94}, 244301 (2005).

\bibitem{LLP03} S. Lepri, R. Livi, A. Politi, 
Phys. Rev. E {\bf 68}, 067102 (2003).

\bibitem{2d} A. Lippi and R. Livi, 
J. Stat. Phys. {\bf 100}, 1147 (2000);
P. Grassberger and L. Yang, cond-mat/0204247; 
L. Delfini {\it et al.} J. Stat. Mech. P05006 (2005).
 
\bibitem{SS97} J. Scheipers, W. Schirmacher, Z. Phys. B \textbf{103}, 547
(1997).

\bibitem{KT} R. Kubo, M. Toda, and N. Hashitsume, {\it Statistical Physics II};
Springer Series in Solid State Sciences, Vol. 31, Springer, Berlin, 1991.

\bibitem{LB94} S. W. Lovesey, E. Balcar, J. Phys. Condens. Matter {\bf 6}
1253 (1994).

\bibitem{adiff} R. Metzler and J. Klafter, Phys. Rep. {\bf 339} 1 (2000).

\bibitem{maina} F. Mainardi, Chaos, Solitons and Fractals \textbf{7} 1461
(1996).

\bibitem{nota} The function $E_\mu$ interpolates between stretched 
exponential and power-law decays (Ref.~\cite{adiff}, App. B). 
The asymptotic form for small arguments yield the same form for 
$C$ as given in the first of Eqs.~(\ref{smallx}).

\bibitem{futuro} L. Delfini, S. Lepri, R. Livi and A. Politi, unpublished.
A similar conclusion has been drawn independently by G. R. Lee-Dadswell et
al. Phys. Rev. E {\bf 72} 031202 (2005).

\end{thebibliography}
\end{document}